\documentclass[aps,showpacs,manuscript,12pt]{revtex4}
\usepackage{amssymb}
\usepackage{amsmath}
\usepackage{graphicx}

\begin{document}
\title{\textbf{Exact pressure evolution equation for incompressible fluids$^{\S }$ }}
\author{M. Tessarotto$^{a,b}$, M. Ellero$^{c}$, N. Aslan$^{d}$, M. Mond$^{e}$
and P. Nicolini $^{a,b}$} \affiliation{ $^{a}$ Department of
Mathematics and Informatics, University of Trieste, Italy,
$^{b}$Consortium of Magneto-fluid-dynamics, University of Trieste,
Italy, $^{c}$Technical University of Munich, Munich, Germany,
$^{d}$Yeditepe University, Kayisdagi, Istanbul, Turkey,
$^{e}$Ben-Gurion University of the Negev, Beer-Sheeva, Israel}
\begin{abstract}
An important aspect of computational fluid dynamics is related to
the determination of the fluid pressure in isothermal
incompressible fluids. In particular this concerns the
construction of an exact evolution equation for the fluid pressure
which replaces the Poisson equation and yields an algorithm which
is a Poisson solver, i.e., it permits to time-advance
exactly the same fluid pressure \textit{without solving the Poisson equation}%
. In fact, the incompressible Navier-Stokes equations represent a
mixture of hyperbolic and elliptic pde's, which are extremely hard
to study both analytically and numerically. In this paper we
intend to show that an exact solution to this problem can be
achieved adopting the approach based on inverse kinetic theory
(IKT) recently developed for incompressible fluids by Ellero and
Tessarotto (2004-2007). In particular we intend to prove that the
evolution of the fluid fields can be achieved by means of a
suitable dynamical system, to be identified with the so-called
Navier-Stokes (N-S) dynamical system. As a consequence it is found
that the fluid pressure obeys a well-defined evolution equation.
The result appears relevant for the construction of Lagrangian
approaches to fluid dynamics.
\end{abstract}
\date{\today }
\maketitle

%$^{\star }$} \url{http://cmfd.univ.trieste.it}.

%%%%%%%%%%%%%%%%%%%%%%%%%%%%%%%%%%%%%%%%%%%%
%% MAINMATTER
%%%%%%%%%%%%%%%%%%%%%%%%%%%%%%%%%%%%%%%%%%%%

\section{Introduction}

The dynamics of incompressible fluids has been around for almost
two centuries. Nevertheless, basic issues concerning its
foundations still remain unanswered. A significant aspect
concerns, in particular, the determination of the fluid pressure
in isothermal incompressible fluids and the construction of
algorithms with permit to time-advance the same fluid pressure. In
fact, the incompressible Navier-Stokes (NS) equations (INSE),
describing a fluid characterized by infinite sound speed,
represent a mixture of hyperbolic and elliptic pde's, which are
extremely hard to study both analytically and numerically. These
difficulties have motivated in the past the search of possible
alternative numerical approaches which permit to advance in time
the fluid pressure without actually solving numerically the
Poisson equation \cite{Harlow,Chorin1967,Turkel99,Succi}. This
refers, especially, to the search of approximate evolution
equations for the fluid pressure based on relaxation-type schemes.
These equations are typically characterized by having finite sound
speed, weak compressibility and, in particular, in the case of
kinetic approaches, also by a low Mach number. All of these
approaches are asymptotic, i.e., they depend on small parameters.
Nevertheless, they are well know to lead to useful numerical
schemes. Notable examples are provided, for example, by the Chorin
artificial compressibility method (ACM) \cite{Chorin1967} and the
lattice Boltzmann approach developed in Ref. \cite{Lee2003}.
However, the interesting question arises whether there exists
actually \emph{an evolution
equation for the fluid pressure} which is \emph{exactly} \emph{equivalent }%
to the Poisson equation (i.e., is a Poisson solver) and does not
require any violation of the fluid equations (in particular the
condition of incompressibility) nor the assumption of low Mach
number. If it exists it could therefore be used, in principle, to
determine improved numerical schemes which permit the evaluation
of the fluid pressure without solving explicitly the Poisson
equation. \ This is a obviously a significant matter-of-principle
issue which should be resolved, even before attempting any
specific solution strategy of this type. The search of an exact
pressure-evolution equation, besides being a still unsolved
mathematical problem, is potentially relevant for several reasons;
in particular: a) the\ proliferation of numerical algorithms in
computational fluid dynamics which reproduce the behavior of
incompressible fluids only in an asymptotic sense (see below); b)
the possible verification of conjectures involving the validity of
appropriate equations of state for the fluid pressure; c)
comparisons with previous CFD methods; d) the identification of
mathematical models which do not involve relaxation-type and/or
low effective-Mach number schemes [to avoid solving the Poisson
equation] and, more generally, the investigation of possible
non-asymptotic phase-space models, i.e., which do not involve
infinitesimal parameters. \ Other possible motivations are, of
course, related to the ongoing quest for efficient numerical
solution methods to be applied for the construction of \ the fluid
fields\ $\left\{ \rho ,\mathbf{V,}p\right\} ,$ solutions of the
initial and boundary-value problem associated to the INSE. The
purpose of this paper is to answer this important question. In
particular, we intend to prove \emph{that the pressure evolution
equation actually exists for incompressible fluids, and can be
constructed explicitly. }The solution of the problem is reached by
adopting a phase-space descriptions of hydrodynamic equations
based on the continuous inverse kinetic theory (IKT) recently
pointed out by Ellero and Tessarotto \cite{Ellero2004,Ellero2005},
in which all the fluid fields, including the fluid pressure
$p(\mathbf{r},t),$ are represented as prescribed momenta of the
kinetic distribution function, solution of an appropriate kinetic
equation. An inverse kinetic theory of this type yields, by
definition, \emph{an exact Navier-Stokes (and also a Poisson)
solver.} The dynamical system (denoted as \emph{Navier-Stokes
dynamical system}), which advances in time the kinetic
distribution function, determines also uniquely the time-evolution
of the fluid fields. As a consequence, it is found the fluid
pressure obeys an evolution equation which is non-asymptotic,
i.e., it is satisfied for arbitrary values of the relevant
physical parameters. This permits to investigate, in particular,
its asymptotic behavior for infinitesimal Mach numbers. We intend
to show that, in validity of suitable asymptotic conditions, it
can be approximated by the Chorin pressure evolution equation.
This result is interesting because it shows that Chorin scheme can
be viewed as an asymptotic approximation to the exact solution
method here presented. \ \ The plan of the presentation is as
follows. In Sec.2 a review of previous numerical solution methods
adopted in CFD for the Poisson equation is presented. \ In Sec. 3
the phase-space approach based on inverse kinetic theory is
recalled. This is realized by constructing the Navier-Stokes
dynamical system which advances in time the kinetic distribution
function. Representing the fluid fields in terms of suitable
velocity-space moments of the kinetic distributions function,
their time evolution is uniquely determined. This result is
invoked in Sec. 4 to construct an exact evolution equation for the
fluid pressure. The equation is proven to hold for arbitrary
Lagrangian trajectories determined by the Navier-Stokes dynamical
system. The physical implications of the equations are discussed
in Sec.5 and its possible asymptotic approximations are discussed.
As a result, it is shown in particular that under suitable
assumptions the pressure evolution equation can be approximated by
the Chorin pressure evolution equation. Finally, the conclusions
are summarized in Sec.6.

\section{Mathematical setting\ - Search of alternative Poisson solvers}

For definiteness, it is convenient to recall that this is defined
by the N-S
and isochoricity equations%
\begin{eqnarray}
&&\left. N\mathbf{V}=\mathbf{0},\right.  \label{INSE-2} \\
&&\left. \nabla \cdot \mathbf{V}=0,\right.  \label{INSE-4}
\end{eqnarray}%
while the mass density $\rho $ satisfies the incompressibility condition $%
\rho (\mathbf{r,}t)=\rho _{o}$, with $\rho _{o}$ a constant mass
density and
both $\rho $, $p$ non-negative. Here $N$ is the N-S operator $N\mathbf{%
V\equiv }\rho \frac{D}{Dt}\mathbf{V}+\mathbf{\nabla
}p+\mathbf{f}-\mu \nabla
^{2}\mathbf{V,}$ with $\frac{D}{Dt}=\frac{\partial }{\partial t}+\mathbf{%
V\cdot \nabla }$ the convective derivative, $\mathbf{f}$ denotes a
suitably smooth volume force density acting on the fluid element
and $\mu >0$ is the constant fluid viscosity. Equations
(\ref{INSE-2})-(\ref{INSE-4}) are
assumed to admit strong solutions in an open set $\Omega \times I,$ with $%
\Omega \subseteq \mathbf{R}^{3}$ the configurations space (defined
as the
subset of $\mathbf{R}^{3}$ where $\rho (\mathbf{r,}t\mathbf{)}>0$) and $%
I\subset \mathbf{R}$ a possibly bounded time interval. By assumption $%
\left\{ \rho ,\mathbf{V,}p\right\} $ are continuous in the closure $%
\overline{\Omega }.$ Hence if in $\Omega \times I,$ $\mathbf{f}$
is at least
$C^{(1,0)}(\Omega \times I),$ it follows necessarily that $\left\{ \rho ,%
\mathbf{V,}p\right\} $ must be at least $C^{(2,1)}(\Omega \times
I),$ while the fluid pressure and velocity must satisfy
respectively the Poisson and energy equations $\nabla
^{2}p=\mathbf{-\triangledown \cdot f-}\rho \nabla
\mathbf{\cdot }\left( \mathbf{V\cdot \nabla V}\right) $ and $\mathbf{V\cdot }%
N\mathbf{V}=0.$ It is well known that the choice of the Poisson
solver results important in numerical simulations, since its
efficient numerical solution depends critically on the number of
modes or mesh points used for its discretization
\cite{Temam1983,Fletcher1997}. In turbulent flows this number can
become so large to effectively limit the size of direct numerical
simulations (DNS) \cite{Ellero2004}. This phenomenon may be
worsened by the algorithmic complexity of the numerical solution
methods adopted for the Poisson equation. For this reason
previously several alternative approaches have been devised which
permit to advance in time the fluid pressure without actually
solving numerically the Poisson equation. Some of these methods
are \emph{asymptotic,} i.e., to advance in time the fluid pressure
they replace the exact Poisson equation with suitable algorithms
or equations which hold only in an asymptotic sense (neglecting
suitably small corrections), others are \emph{exact solvers,}
i.e., provide in principle rigorous solutions of INSE (and Poisson
equation). The first category includes the pressure-based method
(PBM) \cite{Harlow}, the Chorin artificial compressibility method
(ACM) \cite{Chorin1967}, the so-called preconditioning techniques \cite%
{Turkel99}, all based on ACM, and kinetic approaches, of which a
notable example is provided by the so-called Lattice-Boltzmann
(L-B) methods (for a review see for example Ref.\cite{Succi} and
references therein indicated). PBM is an iterative approach and
one of the most widely used for incompressible flows. Its basic
idea is to formulate a Poisson equation for pressure corrections,
and then to update the pressure and velocity fields until the
isochoricity condition (\ref{INSE-4}) is satisfied in a suitable
asymptotic sense. The ACM approach and the related preconditioning
techniques, instead, are obtained by replacing the Poisson and N-S
equations with suitable parameter-dependent evolution equations,
assuming that the fluid fields depend on a fictitious pseudo-time
variable $\tau $. In dimensionless form the evolution equation for
the pressure becomes in such a case
\begin{equation}
\varepsilon ^{2}\frac{\partial }{\partial \tau }p+\nabla \cdot
\mathbf{V=}0, \label{Chorin pressure equation}
\end{equation}%
where $\varepsilon ^{2}>0$ is an infinitesimal parameter.
Manifestly this equation recovers only asymptotically, i.e., for
$\varepsilon ^{2}\rightarrow 0,$ the exact isochoricity condition
(\ref{INSE-4}). Introducing the fast variable $\overline{\tau
}\equiv \tau /\varepsilon
^{2}, $ this implies that the fluid fields must be of the form \ $\mathbf{V}(%
\mathbf{r},t,\overline{\tau }),p(\mathbf{r},t,\overline{\tau })$
and should be assumed suitable smooth functions of $\overline{\tau
}$.$\ $Therefore, for prescribed finite values of $\varepsilon
^{2}$ ( to be assumed suitably small), this equation permits to
obtain also an \emph{asymptotic estimate}
for the fluid pressure $p(\mathbf{r},t)$. This is expressed by the equation $%
p(\mathbf{r},t)=\lim_{\overline{\tau }\rightarrow \infty }p(\mathbf{r},t,%
\overline{\tau })\cong p(\mathbf{r},t,\overline{\tau }=0)-\int\limits_{0}^{%
\overline{\tau }^{\ast }}d\overline{\tau }^{\prime }\nabla \cdot \mathbf{V}(%
\mathbf{r},t,\overline{\tau }^{\prime }),$ where $\overline{\tau
}^{\ast }>>1 $ is suitably defined and
$p(\mathbf{r},t,\overline{\tau }=0)$ denotes some initial estimate
for the fluid pressure. Several implementations on the
Chorin algorithm are known in the literature (see for example Refs.\cite%
{Housman2004,Turkel,Gaitonde}). \ Customary L-B methods are
asymptotic too since they recover INSE only in an approximate
sense; moreover typically they rely on the introduction of an
equation of state for the fluid pressure, for example, the
equation of state of an ideal gas, or more refined models based on
so-called non-ideal fluids \cite{Shi2006}. This assumption,
however, generally requires that the effective Mach-number
characterizing the L-B approach, defined by the ratio
$M^{eff}=V^{\sup }/c$
(with $c$ denoting the discretized velocity of the test particles and $%
V^{\sup \text{ }}$the sup of the velocity field at time $t$),\
must result suitably small. As a consequence, in typical L-B
approaches the fluid pressure can only be estimated
asymptotically. However, there are other numerical approaches
which in principle provide exact Poisson solvers. These include
the so-called spectral methods in which the fluid fields are
expanded in terms of suitable basis functions. Significant
examples are the pure spectral Galerkin and Fourier methods
\cite{Boyd} as well as the nonlinear Galerkin method
\cite{Temam1990}, which are typically adopted for large-scale
turbulence simulations. In these methods the construction of
solution of the Poisson equation is obtained analytically.
However, the series-representation of the fluid fields makes
difficult the investigation of the qualitative properties of the
solutions, such - for example - the search of a possible equation
of state or an evolution equation for the fluid pressure.

\section{Inverse kinetic theory approach to INSE}

Another approach which provides in principle an exact Poisson
solver is the one proposed by Ellero and Tessarotto
\cite{Ellero2004,Ellero2005}, based on an \emph{inverse kinetic
theory} for INSE. This approach, recently applied also to quantum
hydrodynamic equations \cite{Piero}, permits to represent the
fluid fields as moments of a suitably smooth kinetic distribution
function {$f(\mathbf{x},t)$ which obeys an appropriate Vlasov-type
inverse
kinetic equation (IKE):}%
\begin{equation}
\frac{\partial }{\partial t}f+\frac{\partial }{\partial \mathbf{x}}\cdot (%
\mathbf{X}f)=0.  \label{inverse kinetic eq}
\end{equation}%
Here $\mathbf{X(x},t)\equiv \left\{ \mathbf{v,F}\right\} $ and $\mathbf{x}=(%
\mathbf{r,v)\in }\Gamma \subseteq \overline{\Omega }\times
\mathbf{R}^{3}$ is the state vector generated by the vector field
$\mathbf{X,v}$ is the kinetic velocity, while
$\mathbf{F}(\mathbf{x,}t)$ is an appropriate
mean-field force obtained in Ref.\cite{Ellero2005}. \ In Refs. \cite%
{Tessarotto2006,Tessarotto2006b}, it has been proven that. under
suitable assumptions, $\mathbf{F}(\mathbf{x,}t)$ can be uniquely
prescribed. This
implies that t{he time evolution of the kinetic distribution function, }$%
T_{t,t_{o}}f(\mathbf{x}_{o})${\ }${=f(\mathbf{x}(t),t),}$ {is
determined by the finite-dimensional classical dynamical system
associated to the vector
field }$\mathbf{X,}$ i.e.,%
\begin{eqnarray}
&&\left. \frac{d}{dt}\mathbf{x}=\mathbf{X}(\mathbf{x},t)\right.
\label{N-S dynamical system} \\
&&\left. \mathbf{x(}t_{o})=\mathbf{x}_{o}\right.  \notag
\end{eqnarray}%
{\ (\emph{N-S dynamical system}) which must hold for arbitrary
initial conditions
}$\mathbf{x}_{o}=(\mathbf{r}_{o}\mathbf{,v}_{o})\in \Gamma .${\
It follows that the solution of (\ref{N-S dynamical system}), }$\mathbf{x}%
(t)=T_{t,t_{o}}\mathbf{x}_{o},$ which defines the N-S evolution operator $%
T_{t,t_{o}},$ determines uniquely a set of curves $\left\{ \mathbf{x}%
(t)\right\} \equiv \left\{ \mathbf{x}(t),\forall t\in I\right\} _{\mathbf{x}%
_{o}}$ belonging to the phase space $\Gamma $ which can be
interpreted as \emph{phase-space Lagrangian trajectories}
associated to a set of fictitious \emph{"test" particles}. The
projections of these trajectories onto the configuration space,
denoted as \emph{configuration-space} \emph{Lagrangian
trajectories,} are defined by the curves $\left\{
\mathbf{r}(t)\right\} \equiv \left\{ \mathbf{r}(t)\equiv
T_{t,t_{o}}\mathbf{r}_{o},\forall t\in I\right\}
_{\mathbf{x}_{o}}.$ By varying their initial conditions, in
particular $\mathbf{r}_{o}\in \Omega ,$ the curves $\left\{ \mathbf{r}%
(t)\right\} $ can span, by continuity, the whole set
$\overline{\Omega }.$ IKT determines uniquely the fluid fields
expressed via suitable moments of the kinetic distribution
function. The time evolution of the fluid fields requires the
construction of an infinite set of configuration-space Lagrangian
trajectories, each one corresponding to a different initial
position $\mathbf{r}_{o}\in \Omega $. Hence -as expected - INSE is
actually reduced to an infinite set of ode's, represented by the
initial value
problem {(\ref{N-S dynamical system}) with a suitable infinite set of }%
initial conditions. In particular, it follows that the fluid pressure $p(%
\mathbf{r},t)$ is defined by
$p(\mathbf{r},t)=p_{1}(\mathbf{r},t)-p_{o}(t)$
(to be regarded as a{\ \emph{constitutive equation }}for{\emph{\ }} $p(%
\mathbf{r},t)$), where {$p_{1}(\mathbf{r},t)$} is the {\emph{kinetic pressure%
} $p_{1}(\mathbf{r},t)=\int dv\frac{1}{3}u^{2}f(\mathbf{x},t),$
}$p_{o}$\ is denoted as \emph{\ reduced pressure, }while{\
}$\mathbf{u}$ is the relative velocity $\mathbf{u}\mathbf{\equiv
}\mathbf{v}-\mathbf{V}(\mathbf{r,}t).$
Both $p_{o}(t)$ and {$p_{1}(\mathbf{r},t)$} are strictly positive, while $%
p_{o}(t)$ in $\overline{\Omega }\times I$ is subject to the constraint $%
p_{1}(\mathbf{r},t)-p_{o}(t)\geq 0.$ In particular the reduced pressure $%
p_{o}(t)$ is an arbitrary function of time, subject to the only
requirements of suitably smoothness and strict positivity. A key
aspect of IKT is that, by construction, a particular solution of
IKE is provided by the local
Maxwellian distribution $f_{M}(\mathbf{x,}t;\mathbf{V,}p_{1})=\frac{\rho _{o}%
}{\left( \pi \right) ^{\frac{3}{2}}v_{th}^{3}}\exp \left\{
-Y^{2}\right\} $ [where $Y^{2}=\frac{\mathbf{u}^{2}}{vth^{2}}$ and
$v_{th}^{2}=2p_{1}/\rho
_{o}$]. In such a case, the vector field $\mathbf{F}$ reads $\mathbf{F(r,v,}%
t)=\mathbf{a}-\frac{1}{\rho }N_{0}\mathbf{V}+\frac{\mathbf{u}}{2}A_{0}p_{1}+%
\frac{1}{\rho }\nabla p\left\{
\frac{\mathcal{E}}{p_{1}}-\frac{3}{2}\right\} $ (\emph{equation
for} $\mathbf{F}$)$\mathbf{,}$ where $\mathbf{a}$ denotes
the convective term $\mathbf{a=}\frac{1}{2}\mathbf{u}\cdot \nabla \mathbf{V+}%
\frac{1}{2}\nabla \mathbf{V\cdot u,}$ $\mathcal{E}$ is the
relative kinetic
energy density $\mathcal{E}\mathcal{=}\rho u^{2}/2,$ while $N_{0}$\textbf{\ }%
and\textbf{\ }$A_{0}$ are the differential operators $N_{0}\mathbf{V}\equiv -%
\mathbf{f(r,V,}t)+\mu \nabla ^{2}\mathbf{V}$ and $A_{0}p_{1}\mathbf{(r,}%
t)\equiv \frac{1}{p_{1}}\left[ \frac{\partial }{\partial
t}p_{1}+\nabla \cdot \left( \mathbf{V}p_{1}\right) \right] $. For
an arbitrary and suitably smooth distribution function
$f(\mathbf{x,}t),$ the form of the vector field
$\mathbf{F}$ satisfying these hypotheses has been given in Refs. \cite%
{Ellero2005,Tessarotto2006}.

\section{Construction of an exact pressure evolution equation}

An interesting issue is related to the consequences of the
constitutive equation for $p$ and the N-S dynamical system
generated by the initial value-problem (\ref{N-S dynamical
system}). \ In this Letter we intend to prove that the fluid
pressure $p(\mathbf{r},t)$ obeys an exact partial-differential
equation which uniquely determines is time evolution. This is
obtained by evaluating its Lagrangian derivative along an
arbitrary configuration-space Lagrangian trajectory $\left\{
\mathbf{r}(t)\right\} $ generated by the N-S dynamical system. The
result can be stated as follows.

{\bf Theorem - Pressure evolution equation}\\
\textit{Assuming that the initial-boundary value problem
associated to INSE
admits a suitably strong solution }$\left\{ \rho ,\mathbf{V,}p\right\} $%
\textit{\ in the set }$\Omega \times I,$\textit{\ the following
statements hold:}

\textit{A) If }$\mathbf{x}(t)$\emph{\ is a particular solution of} \textit{%
Eq.} \textit{\ (\ref{N-S dynamical system}) which holds for arbitrary }$%
\mathbf{r}(t)\in \Omega $ \textit{and} $t\in I,$ \textit{along
each
phase-space Lagrangian trajectory }$\left\{ \mathbf{x}(t)\right\} $ \textit{%
defined by Eq.} \textit{\ (\ref{N-S dynamical system}) the}
\textit{scalar field} $\xi (\mathbf{r},t)\equiv $
$\mathcal{E}/p_{1}$ \textit{obeys the
exact evolution equation}%
\begin{equation}
\frac{d}{dt}\xi =-\frac{1}{2}\mathbf{u\cdot }\nabla \ln p_{1}
\label{evolution equation -1}
\end{equation}%
\textit{which holds for arbitrary} \textit{\ initial conditions} $\mathbf{x}%
_{o}{}=\mathbf{(r}_{o}\mathbf{,v}_{o}\mathbf{),}$ \textit{and} $\xi _{o}=%
\frac{\rho u_{o}^{2}}{2p_{1}(\mathbf{r}_{o},t_{o})},$ \textit{with} $\mathbf{%
u}_{o}\equiv \mathbf{v}_{o}-\mathbf{V}(\mathbf{r}_{o},t_{o}).$
\textit{Here is }$\frac{d}{dt}$\textit{\ the Lagrangian derivative
}$\frac{d}{dt}\equiv
\frac{\partial }{\partial t}+\mathbf{v}\cdot \nabla +\mathbf{F}\cdot \frac{%
\partial }{\partial \mathbf{v}},$\textit{\ }$\xi (\mathbf{r},t),$ \textit{%
while all quantities (}$\mathbf{u},E$\textit{\ and
}$p_{1})$\textit{\ are
evaluated along an arbitrary phase-space trajectory \ }$\left\{ \mathbf{x}%
(t)\right\} .$

\textit{B) Vice versa, if the solutions} $\mathbf{x}(t)\mathbf{=}(\mathbf{r}%
(t)\mathbf{,\mathbf{v}(}t))$ \textit{and }$\xi (t)$\textit{\ of Eqs.(\ref%
{N-S dynamical system}), (\ref{evolution equation -1}) are known
for
arbitrary initial conditions} $\mathbf{x}_{o}{}=\mathbf{(r}_{o}\mathbf{,v}%
_{o}\mathbf{),}$ $\mathbf{u}_{o}\equiv \mathbf{v}_{o}-\mathbf{V}(\mathbf{r}%
_{o},t_{o})$ \textit{and} $\xi _{o}=\frac{\rho u_{o}^{2}}{2p_{1}(\mathbf{r}%
_{o},t_{o})})$\textit{\ and for all }$(\mathbf{r},t)\in $$\Omega
\times I,$
\textit{\ it follows necessarily that in }$\Omega \times I,$ $\left\{ \rho ,%
\mathbf{V,}p\right\} $ \textit{satisfy identically INSE.}

PROOF

Let us first prove statement A), namely that INSE and the N-S
dynamical system imply necessarily the validity of \
Eq.(\ref{evolution equation -1}). For this purpose we first notice
that by construction Eq.(\ref{N-S dynamical system}) admits a
unique solution $\mathbf{x}(t)$ for arbitrary initial conditions
$\mathbf{x}_{o}=(\mathbf{r}_{o}\mathbf{,v}_{o})\in \Gamma ,$ while
the same equation can also be expressed in terms of the relative
velocity $\mathbf{u\mathbf{=v-V}}(\mathbf{\mathbf{r},}t)$. This
yields

\begin{equation}
\frac{d}{dt}\mathbf{u=F-}\frac{D\mathbf{V}(\mathbf{r,}t\mathbf{)}}{Dt}-%
\mathbf{u\cdot \nabla V}(\mathbf{r,}t\mathbf{)}  \label{eq-10}
\end{equation}%
Upon invoking the N-S equation (\ref{INSE-2}) and by taking the
scalar
product of Eq.(\ref{eq-10}) by $\rho \mathbf{u}$, this equation implies $%
\frac{d}{dt}\mathcal{E}\mathcal{=}\mathbf{u\cdot }\nabla p_{1}\left\{ \frac{%
\mathcal{E}}{p_{1}}-\frac{1}{2}\right\}
+\frac{\mathcal{E}}{p_{1}}\left[
\frac{\partial }{\partial t}p_{1}+\nabla \cdot \left( \mathbf{V}p_{1}\right) %
\right] ,$ which finally gives
\begin{equation}
\frac{d}{dt}\xi \equiv \frac{\partial }{\partial t}\xi
+\mathbf{v\cdot
\nabla }\xi +\mathbf{F\cdot }\frac{\partial }{\partial \mathbf{v}}\xi =-%
\frac{1}{2p_{1}}\mathbf{u\cdot }\nabla
p_{1}+\frac{\mathcal{E}}{p_{1}}\nabla \cdot \mathbf{V}.
\label{evolution equation for p}
\end{equation}%
As a consequence of the isochoricity condition (\ref{INSE-4}) this
equation reduces identically (i.e., for arbitrary initial
conditions for the dynamical system) to Eq.(\ref{evolution
equation -1}). \ \ B) Vice versa,
let us assume that the solutions $\mathbf{x}(t)\mathbf{=}(\mathbf{r}(t)%
\mathbf{,\mathbf{v}(}t))$ and $\xi (t)$ of Eqs.(\ref{N-S dynamical
system}),
(\ref{evolution equation -1}) are known for arbitrary initial conditions $%
\mathbf{x}_{o}\in \Gamma $ and $\xi _{o}=\frac{\rho u_{o}^{2}}{2p_{1}(%
\mathbf{r}_{o},t_{o})}.$ In this case it follows the fluid fields
necessarily must satisfy INSE in the whole set $\Omega \times I.$
It
suffices, in fact, to notice that by assumption the evolution operator $%
T_{t,t_{o}}$ is known. This permits to determine uniquely the
kinetic
distribution function at time $t,$ which reads \cite{Ellero2005} $f(\mathbf{x%
}(t),t)=f(\mathbf{x}_{o},t_{o})/J(\mathbf{x}(t),t),$ where $J(\mathbf{x}%
(t),t)$ is the Jacobian of the flow $\mathbf{x}_{o}\rightarrow
\mathbf{x}(t).
$ Hence, also its moments are uniquely prescribed, including both $\mathbf{%
V(r,}t)$ and \ $p(\mathbf{r,}t),$ in such a way that they result at least $%
C^{(2,1)}(\Omega \times I).$ The inverse kinetic equation
(\ref{inverse kinetic eq}), thanks to the special form of
$\mathbf{F,}$ as given by its
definition, ensures that the N-S equation is satisfied identically in $%
\Omega \times I$ \cite{Ellero2005}. Moreover, since Eqs. (\ref{eq-10}) and (%
\ref{evolution equation for p}) are by assumption fulfilled
simultaneously, it follows that both the isochoricity condition
(\ref{INSE-4}) and the Poisson equation must be satisfied too in
$\Omega \times I.$ This completes the proof.

Let us analyze the consequences of the theorem.\ \ If the fluid
velocity is assumed to satisfy both the N-S equation and
isochoricity condition, the mass density satisfies the
incompressibility condition, while $\left\{ \mathbf{x}(t)\right\}
$ is an arbitrary trajectory of the N-S dynamical system, it
follows that Eq.(\ref{evolution equation -1}) determines uniquely
the time-advancement of the fluid pressure. Hence, it provides an
evolution equation for the fluid pressure, which by definition is
equivalent to the Poisson equation [or to the equivalent problem
obtained imposing, incompressibility, \ isochoricity and
Navier-Stokes equations]. This equation can in principle be used
to determine the fluid pressure at an arbitrary position
$\mathbf{r}\in \Omega .$ However, since any given position can be
reached by infinite phase-space (and also configuration-space)
Lagrangian trajectories, it is sufficient to sample the
configuration space by a suitable subset of Lagrangian
trajectories (test particles), obtained by prescribing the initial
condition $\mathbf{x}_{o}$. The physical interpretation of the
pressure evolution equation is elementary: it prescribes the
Lagrangian time derivative of the fluid
pressure, which is defined in the frame which is locally at the position $%
\mathbf{r}(t)$ is co-moving with the test particle velocity $\mathbf{\mathbf{%
v}(}t).$ Let us now exploit the arbitrariness in the definition of
the reduced pressure $p_{o}(t)$ and of the initial kinetic
velocity $\mathbf{v}_{o}.$ Since IKT holds for arbitrary values of
both parameters, this means that the
dimensionless ratios $M_{V}=V/\left\vert \mathbf{v}_{o}\right\vert $ and $%
M_{p}=p/p_{o},$ to be denoted as \emph{velocity} and
\emph{pressure effective Mach numbers}, remain essentially free. \
As a consequence, it is possible: a) to construct asymptotic
solutions of the pressure evolution
equation based on low effective-Mach numbers expansions, i.e., for which $%
M_{V},$ $M_{p}\ll 1$ and moreover also b) to determine asymptotic
approximations to Eq.(\ref{evolution equation -1}). In particular,
it is immediate to show that the pressure evolution equation
admits the Chorin pressure evolution equation (\ref{Chorin
pressure equation}) as its leading-order asymptotic approximation.
\ In fact, invoking the asymptotic ordering $M_{V}\sim O(\delta
),$ there results in the leading-order approximation
$\frac{d}{dt}\ln \mathcal{E\cong }\mathbf{u\cdot }\nabla \ln
p_{1}\left[ 1+o(\delta )\right] .$ This implies $\frac{d}{dt}\xi \cong \frac{%
1}{p_{1}}\mathbf{u\cdot }\nabla \ln p_{1}\left[ 1+o(\delta
)\right] .-\xi \frac{d}{dt}\ln p_{1}.$ Invoking Eq.(\ref{evolution
equation for p}) and requiring also $M_{p}\sim O(\delta ),$ with
$p_{o}=const.,$ there results, again to leading order in $o(\delta
)$
\begin{equation}
\frac{1}{p_{o}}\frac{d}{dt}p+\nabla \cdot \mathbf{V\cong 0.}
\label{asymptotic pressure equation}
\end{equation}%
This equation reduces to the Chorin's Eq.(\ref{Chorin pressure
equation}),
if the representation $\frac{1}{p_{o}}\frac{d}{dt}p\equiv \varepsilon ^{2}%
\frac{\partial }{\partial \tau }p$ is adopted. Hence, in this
sense, Chorin's pressure equation may simply be viewed as an
asymptotic approximation to the exact pressure equation
(\ref{evolution equation -1}).
Denoting $p_{o}\equiv \rho _{o}c^{2},$ Eq.(\ref{asymptotic pressure equation}%
), together with the Navier-Stokes equation (\ref{INSE-2}),
describes a weakly compressible fluid characterized by a finite
sound velocity $c$. \ This result suggests, however, also the
possibility of constructing more accurate approximations for the
pressure equation, which are higher-order in $\delta $ and have
prescribed accuracy with respect to $\delta $.

\section{Conclusions}
The mere fact that an approximate evolution equation for the
pressure can be obtained for incompressible fluids is not new.
Examples of efficient Poisson solvers of this type are well known
\cite{Chorin1967,Lee2003}. These equations, however, are usually
asymptotic, since the limit of incompressibility is achieved only
for infinite sound speeds. Contrary to the common wisdom that for
incompressible fluids it can only be achieved adopting
relaxation-type models, in this paper we have proven that an exact
local evolution equation can be obtained without modifying the
incompressible Navier-Stokes equations, in particular without
introducing the assumption of weak compressibility (and low Mach
number). The proof has
been reached by adopting an inverse kinetic theory \cite%
{Ellero2004,Ellero2005} which permits the identification of the
(Navier-Stokes) dynamical system and of the corresponding
evolution operator which advances in time the kinetic distribution
function and the related fluid fields. A remarkable feature of the
pressure evolution equation here obtained is that it is
non-asymptotic and holds for arbitrary phase-space Lagrangian
trajectories generated by the same dynamical system. This makes it
possible, by suitably selecting these trajectories, to construct,
in principle, asymptotic approximations with \emph{prescribed
accuracy} both for the pressure evolution equation and for its
solutions. In particular, we have proven that under appropriate
asymptotic assumptions to leading order (in the relevant
asymptotic parameter) the well known Chorin pressure evolution
equation is recovered. \ These results are important both from the
conceptual viewpoint and for their possible applications in CFD.
In particular, an interesting open problem is related to the
numerical implementation of the pressure equation (\ref{evolution
equation -1}) in CFD schemes. These developments will be the
object of future investigations. \

%% BACKMATTER
%%%%%%%%%%%%%%%%%%%%%%%%%%%%%%%%%%%%%%%%%%%%%%%%

%% If the bibliography is produced without BibTeX comment out the

\section*{Acknowledgments}
Work developed in cooperation with the CMFD Team, Consortium for
Magneto-fluid-dynamics (Trieste University, Trieste, Italy). \
Research developed in the framework of the MIUR (Italian Ministry
of University and Research) PRIN Programme: {\it Modelli della
teoria cinetica matematica nello studio dei sistemi complessi
nelle scienze applicate}. The support (N.A., M.M and M.E.) of ICTP
(International Center for Theoretical Physics, Trieste, Italy),
(N.A. and M.M.) COST Action P17 (EPM, {\it Electromagnetic
Processing of Materials}) and (M.T. and P.N.) GNFM (National Group
of Mathematical Physics) of INDAM (Italian National Institute for
Advanced Mathematics) is warmly acknowledged.

\section*{Notice}
$^{\S }$ contributed paper at RGD26 (Kyoto, Japan, July 2008).
\newpage

%% BACKMATTER
%%%%%%%%%%%%%%%%%%%%%%%%%%%%%%%%%%%%%%%%%%%%%%%%

%% If the bibliography is produced without BibTeX comment out the


\newpage
\begin{thebibliography}{BIBTEX}
\bibitem{Harlow} F. H. Harlow and J. E. Welch, Phys. Fluids \textbf{8}, 2182
(1965).

\bibitem{Chorin1967} A.J. Chorin, J. Comp. Phys. \textbf{2}, 12 (1967).

\bibitem{Turkel99} E. Turkel, Annual Reviews in Fluid Mechanics 1999,
\textbf{31}, 385-416 (1999).

\bibitem{Succi} S. Succi, \textit{The Lattice-Boltzmann Equation for Fluid
Dynamics and Beyond (Numerical Mathematics and Scientific
Computation)}, Oxford Science Publications (2001).

\bibitem{Lee2003} T. Lee and C-L. Lin, Phys.Rev. E \textbf{67}, 056703
(2003).

\bibitem{Ellero2004} M. Tessarotto and M. Ellero, Proc. 24th
RGD, Bari, Italy (July 2004), Ed. M. Capitelli, AIP Conf. Proc.
\textbf{762}, 108 (2005).

\bibitem{Ellero2005} M. Ellero and M. Tessarotto, Physica A \textbf{355},
233 (2005).

\bibitem{Tessarotto2006} M. Tessarotto and M. Ellero, Physica A \textbf{373}%
, 142 (2007); arXiv: physics/0602140.

\bibitem{Tessarotto2006b} M. Tessarotto and M. Ellero, Proc.
25th RGD (International Symposium on Rarefied gas Dynamics, St.
Petersburg, Russia, July 21-28, 2006), Ed. M.S. Ivanov and A.K.
Rebrov (Novosibirsk Publ. House of the Siberian Branch of the
Russian Academy of Sciences), p.1001; arXiv:physics/0611113 (2007)

\bibitem{Temam1983} C. Foias, O.P. Manley, R. Temam and Y.M. Treve,
Phys.Rev. Lett. \textbf{50}, 1031 (1983).

\bibitem{Fletcher1997} C.A.J. Fletcher, \textit{Computational Techniques for
Fluid Dynamics}, Vol.I pag. 190-192, Springer-Verlag, Berlin,
Heidelberg, New York (1997).

\bibitem{Housman2004} J. Housman, C. Kiris and D. Kwark, Comp. Fluid Dyn. J.
\textbf{13}(3), 483 (2004).

\bibitem{Turkel} E. Turkel, Applied Numerical Mathematics \textbf{12}, 257
(1993).

\bibitem{Gaitonde} A.L. Gaitonde, Int. J. Num. Meth. in Eng \textbf{41,}
1153 (1998).

\bibitem{Shi2006} Y. Shi, T. S. Zhao and Z. L. Guo, Phys. Rev.E \textbf{73,}
026704 (2006).

\bibitem{Boyd} J.P. Boyd, \textit{Chebyshev and Fourier Spectral Methods},
DOVER Publications Inc., New York (2000).

\bibitem{Temam1990} F. Jauberteau, C. Rosier and R. Temam, App. Num. Math.
\textbf{6}, 361--370 (1990).

\bibitem{Piero} M. Tessarotto, M. Ellero and P. Nicolini, {\it
Inverse kinetic theory for quantum hydrodynamic equations},
Phys.Rev. A\textbf{75}, 060691 (2007); arXiv:quantum-ph/060691.

\end{thebibliography}
\end{document}